\documentclass[11pt]{article}

\usepackage{fullpage,latexsym}
\usepackage{latexsym}

\def\01{\{0,1\}}

\newcommand{\eps}{\varepsilon} 
\newcommand{\ket}[1]{|#1\rangle}
\newcommand{\bra}[1]{\langle#1|}
\newcommand{\ketbra}[2]{|#1\rangle\langle#2|}
\newcommand{\inp}[2]{\langle{#1}|{#2}\rangle} 
\newcommand{\norm}[1]{\mbox{$\parallel{#1}\parallel$}} 
\newcommand{\Tr}{\mbox{\rm Tr}}
\newtheorem{definition}{Definition}
\newtheorem{theorem}{Theorem}
\newtheorem{lemma}{Lemma}

\newcommand{\ignore}[1]
{}

\newenvironment{proof}
{\noindent {\bf Proof. }}
{{\hfill $\Box$}\\
 \smallskip}

\begin{document}

\begin{titlepage}
\title{Exponential Lower Bound for 2-Query Locally Decodable Codes\\
via a Quantum Argument}
\author{Iordanis Kerenidis\thanks{Supported by DARPA under agreement number F 30602--01-2--0524.}\\ UC Berkeley\\ {\tt jkeren@cs.berkeley.edu}
\and
Ronald de Wolf\thanks{Most of this work was done while a postdoc at UC Berkeley,
supported by Talent grant S 62--565 from the Netherlands Organization 
for Scientific Research (NWO).}\\ CWI, Amsterdam\\ {\tt rdewolf@cwi.nl}
}
\date{}
\maketitle

\begin{abstract}
A locally decodable code encodes $n$-bit strings $x$ in $m$-bit codewords $C(x)$, 
in such a way that one can recover any bit $x_i$ from a corrupted codeword
by querying only a few bits of that word.  
We use a {\em quantum\/} argument to prove that LDCs with 2 classical 
queries need exponential length: $m=2^{\Omega(n)} $.  
Previously this was known only for linear codes (Goldreich et al.~02).    
Our proof shows that a 2-query LDC can be decoded with only 1 quantum query,
and then proves an exponential lower bound for such 1-query locally
quantum-decodable codes.  We also show that $q$ quantum queries
allow more succinct LDCs than the best known LDCs with $q$ classical
queries.  Finally, we give new classical lower bounds and quantum
upper bounds for the setting of private information retrieval.
In particular, we exhibit a quantum 2-server PIR scheme with
$O(n^{3/10})$ qubits of communication, improving upon the $O(n^{1/3})$
bits of communication of the best known classical 2-server PIR.\\[3mm]
\end{abstract}
\thispagestyle{empty}
\end{titlepage}

\section{Introduction}


Error-correcting codes allow one to encode an $n$-bit string $x$ into
an $m$-bit codeword $C(x)$, in such a way that $x$ can still be recovered
even if the codeword is corrupted in a number of places. 
For example, codewords of length $m=O(n)$ already suffice to recover 
from errors in a constant fraction of the bitpositions of the codeword 
(even in linear time~\cite{sipser&spielman:ec}).
One disadvantage of such ``standard'' error-correction, is that
one usually needs to consider all or most of the (corrupted) codeword
to recover anything about $x$. If one is only interested in recovering
one or a few of the bits of $x$, then more efficient schemes are possible,
so-called locally decodable codes (LDCs).
LDCs allow us to extract small parts of encoded information from a 
corrupted codeword, while looking at (``querying'') only
a few positions of that word.
They have found various applications in complexity theory and cryptography,
such as self-correcting computations, PCPs, 
worst-case to average-case reductions, and private information retrieval.
Informally, LDCs are described as follows:
\begin{quote}
A {\em $(q,\delta,\eps)$-locally decodable code} 
encodes $n$-bit strings $x$ into $m$-bit codewords $C(x)$, 
such that for each $i$, the bit $x_i$ can be recovered with probability
$1/2+\eps$ making only $q$ queries,
even if the codeword is corrupted in $\delta m$ of the bits.
\end{quote}
For example, the Hadamard code is a locally decodable code where 
{\em two} queries are sufficient in order to predict any bit with 
constant advantage, even with a constant fraction of errors. 
The code has $m=2^n$ and $C(x)_j=j\cdot x\mbox{ mod }2$
for all $j\in\01^n$. Recovery from a corrupted codeword $y$ is possible 
by picking a random $j\in\01^n$, querying $y_j$ and $y_{j\oplus e_i}$,
and outputting the XOR of those two bits. If neither bit has been
corrupted, then we output $y_j\oplus y_{j\oplus e_i}=
j\cdot x\oplus(j\oplus e_i)\cdot x=e_i\cdot x=x_i$, as we should.
If $C(x)$ has been corrupted in at most $\delta m$ positions,
then a fraction of at least $1-2\delta$ of all $(j,j\oplus e_i)$ 
pairs of indices is uncorrupted, so the recovery probability 
is at least $1-2\delta$. This is $>1/2$ as long as $\delta<1/4$.
The main drawback of the Hadamard code is its exponential length.

Clearly, we would like both the codeword length $m$ and the number 
of queries $q$ to be small.
The main complexity question about LDCs is how large 
$m$ needs to be, as a function of $n$, $q$, $\delta$, and $\eps$.
For $q=polylog(n)$, Babai et al.~\cite{bfls:checking} showed 
how to achieve length $m=O(n^2)$, for some fixed $\delta,\eps$.
This was subsequently improved to nearly linear length by 
Polishchuk and Spielman~\cite{polishchuk&spielman:holographic}.
Beimel et al.~\cite{bikr:improvedpir} recently improved the best 
known upper bounds for constant $q$ to $m=2^{n^{O(\log\log q/q\log q)}}$,
with some more precise bounds for small $q$.


The study of {\em lower\/} bounds on $m$ was initiated by 
Katz and Trevisan~\cite{katz&trevisan:ldc}.
They proved that for $q=1$, LDCs do not exist if $n$ is larger
than some constant depending on $\delta$ and $\eps$.
For $q\geq 2$, they proved a bound of $m=\Omega(n^{q/(q-1)})$ if the
$q$ queries are made non-adaptively; this bound was generalized 
to the adaptive case by Deshpande et al.~\cite{djklr:ldc}.
This establishes superlinear but at most quadratic lower bounds on 
the length of LDCs with a constant number of queries.
There is still a large gap between the best known upper and lower bounds.
In particular, it is open whether $m=poly(n)$ is achievable with constant $q$.
Recently, Goldreich et al.~\cite{gkst:lowerpir} examined the case $q=2$,
and showed that $m\geq 2^{\delta\eps n/8}$ if $C$ is a {\em linear} code. 
Obata~\cite{obata:ldc} subsequently strengthened the dependence on $\eps$ 
to $m\geq 2^{\Omega(\delta n/(1-2\eps))}$, which is essentially optimal.

Katz and Trevisan, and Goldreich et al.~established a close
connection between locally decodable codes and 
{\em private information retrieval (PIR)\/} schemes. 
In fact, the best known LDCs for constant $q$ are derived from PIR schemes.
A PIR scheme allows a user to extract a bit $x_i$ 
from an $n$-bit database $x$ that is
replicated over some $k\geq 1$ servers, without the server(s) 
learning {\em which\/} $i$ the user wants.
The main complexity measure of a PIR scheme is its communication complexity,
i.e., the sum of the lengths of the queries that the user sends to 
each server, and the length of the servers' answers.
If there is only one server ($k=1$), then privacy can be maintained
by letting the server send the whole $n$-bit database to the user.
This takes $n$ bits of communication and is optimal.
If the database is replicated over $k\geq 2$ servers, then 
smarter protocols are possible. 
Chor et~al.~\cite{cgks:pir} exhibited a 2-server PIR scheme with 
communication complexity $O(n^{1/3})$ and one with $O(n^{1/k})$
for $k>2$. Ambainis~\cite{ambainis:pir} improved the latter 
to $O(n^{1/(2k-1)})$. 
Beimel et al.~\cite{bikr:improvedpir} improved 
the communication complexity to $O(n^{2\log\log k/k\log k})$;
their results improve the previous best bounds for all $k\geq 3$ 
but not for $k=2$.
No general lower bounds better than $\Omega(\log n)$ are known
for PIRs with $k\geq 2$ servers. 
A PIR scheme is {\em linear\/} if for every query the user makes,
the answer bits are  {\em linear combinations\/} of the bits of $x$.
Goldreich et al.~\cite{gkst:lowerpir} proved that linear 2-server 
PIRs with $t$-bit queries and $a$-bit answers where the user looks only 
at $k$ predetermined positions in each answer, require $t=\Omega(n/a^k)$.

\subsection{Our results: Locally decodable codes}

The main result of this paper is an exponential lower bound for
general 2-query LDCs:
\begin{quote}
A $(2,\delta,\eps)$-locally decodable code requires length $m\geq 2^{cn-1}$,
\end{quote}
for $c=1-H(1/2+3\delta\eps/14)$, where $H(\cdot)$ is the binary 
entropy function. This is the first superpolynomial lower bound on 
general LDCs with more than one query.
Our constant $c$ in the exponent is somewhat worse than the ones of 
Goldreich et al.\ and of Obata, but our proof establishes 
the exponential lower bound for {\em all\/} LDCs, not just linear ones.
In the body of the paper we will focus only on codes over 
the {\em binary\/} alphabet. In Appendix~\ref{applarge} we show how 
to extend our result to the case of larger alphabets,
using a classical reduction due to Trevisan.

Our proof introduces one radically new ingredient: {\em quantum\/} computing.
We show that if two classical queries can recover $x_i$ with probability 
$1/2+\eps$, then $x_i$ can also be recovered with probability $1/2+4\eps/7$
using only one quantum query.\footnote{One can't reduce {\em 3\/} classical 
queries to 1 quantum query, because the XOR of 3 bits requires
2 quantum queries.}  
In other words, a $(2,\delta,\eps)$-locally decodable code
is a $(1,\delta,4\eps/7)$-locally {\em quantum}-decodable code.
We then prove an exponential lower bound for 1-query LQDCs by
showing, roughly speaking, that a 1-query LQDC of length $m$
induces a {\em quantum random access code} for $x$ of length $\log m$.
Nayak's~\cite{nayak:qfa} linear lower bound on such codes 
finishes off the proof. For the sake of completeness,
we include a proof of his result in Appendix~\ref{appqrac}.

This lower bound for classical LDCs is one of the very few examples 
where tools from quantum computing enable one to prove {\em new\/} 
results in {\em classical\/} computer science.
We know only a few other examples of this.%
\footnote{The quantum lower bound on the communication complexity of
the inner product function of Cleve et al.~\cite{cdnt:ip} provides
new insight in a classical result, but does not establish a {\em new\/} 
result for classical computer science.}
Radhakrishnan et al.~\cite{rsv:qsetmembership} proved lower bounds
for the set membership data structure that hold for quantum algorithms,
but are in fact stronger than the previous classical lower bounds of Buhrman 
et al.~\cite{bmrv:bitvectors}. 
Sen and Venkatesh did the same for data structures for the predecessor problem
\cite[quant-ph version]{sen&venkatesh:qcellprobe}.
Finally, Klauck et al.~\cite{kntz:qinteraction} proved lower bounds 
for the $k$-round quantum communication complexity of the tree-jumping 
problem that are somewhat stronger than the previous best 
classical lower bounds. 
In these cases, however, the underlying proof techniques
easily yield a classical proof. Our proof seems to be 
more inherently ``quantum'' since there is no classical analog of our 
2-classical-queries-to-1-quantum-query reduction 
(2-query LDCs exist but 1-query LDCs don't).

We also observe that our construction implies the existence 
of 1-query quantum-decodable codes for all $n$.
The Hadamard code is an example of this.
Here the codewords are still classical, but the decoding algorithm is quantum.
As mentioned before, if we only allow one {\em classical\/} query,
then LDCs do not exist for $n$ larger than some constant 
depending on $\delta$ and $\eps$~\cite{katz&trevisan:ldc}.
For larger $q$, it turns out that the best known $(2q,\delta,\eps)$-LDCs,
due to Beimel et al.~\cite{bikr:improvedpir},
are actually $(q,\delta,\eps)$-LQDCs. 
Hence for fixed number of queries $q$, we obtain
LQDCs that are significantly shorter than the best known LDCs.
In particular, Beimel et al.~give a 4-query LDC with length $m=2^{O(n^{3/10})}$
which is a 2-query LQDC.
This is significantly shorter than the $m=2^{\Theta(n)}$ that 
2-query LDCs need.
We summarize the situation in the following table, 
where our contributions are indicated by boldface.

\begin{table}[h]
\centering
\begin{tabular}{|c|c|c|}
\hline
Queries & Length of LDC & Length of LQDC\\ \hline\hline
$q=1$ & don't exist        & {\boldmath $2^{\Theta(n)}$} \\ \hline
$q=2$ & {\boldmath $2^{\Theta(n)}$} & {\boldmath $2^{O(n^{3/10})}$} \\ \hline
$q=3$ & $2^{O(n^{1/2})}$ & {\boldmath $2^{O(n^{1/7})}$} \\ \hline
$q=4$ & $2^{O(n^{3/10})}$ & {\boldmath $2^{O(n^{1/11})}$} \\ \hline
\end{tabular}
\caption{Best known bounds on the length 
of LDCs and LQDCs with $q$ queries}
\end{table}

\subsection{Our results: Private information retrieval}

In the private information retrieval setting,
our techniques allow us to reduce classical 2-server PIR schemes 
with 1-bit answers to quantum 1-server PIRs, 
which in turn can be reduced to a random access code~\cite{nayak:qfa}. 
Thus we obtain an $\Omega(n)$ lower bound on the communication complexity 
for all classical 2-server PIRs with 1-bit answers.
Previously, such a bound was known only for {\em linear\/} PIRs 
(first proven in~\cite[Section~5.2]{cgks:pir} and extended 
to linear PIRs with constant-length answers in~\cite{gkst:lowerpir}).
In Appendix~\ref{applarge} we extend our lower bound to
PIR schemes with larger answers.

Apart from giving new lower bounds for {\em classical\/} PIR,
we can also use our 2-to-1 reduction to obtain {\em quantum\/} PIR
schemes that beat the best known classical PIRs.
In particular, Beimel et al.~\cite[Example~4.2]{bikr:improvedpir} exhibit 
a classical 4-server PIR scheme with 1-bit answers and 
communication complexity $O(n^{3/10})$. We can reduce
this to a quantum 2-server PIR with $O(n^{3/10})$ qubits of
communication. This beats the best known classical 2-server PIR,
which has complexity $O(n^{1/3})$. 
We can similarly give quantum improvements over the best known $k$-server
PIR schemes for $k>2$.
However, this does not constitute a true classical-quantum separation 
in the PIR setting yet, since no good lower bounds are known for classical PIR.
We summarize the best known bounds for classical and quantum PIR below.

\begin{table}[h]
\centering
\begin{tabular}{|c|c|c|}
\hline
Servers & PIR complexity & QPIR complexity\\ \hline\hline
$k=1$ & $\Theta(n)$      & $\Theta(n)$ \\ \hline
$k=2$ & $O(n^{1/3})$     & {\boldmath $O(n^{3/10})$} \\ \hline
$k=3$ & $O(n^{1/5.25})$     & {\boldmath $O(n^{1/7})$} \\ \hline
$k=4$ & $O(n^{1/7.87})$     & {\boldmath $O(n^{1/11})$} \\ \hline
\end{tabular}
\caption{Best known bounds on the communication complexity of
classical and quantum PIR}
\end{table}

\section{Preliminaries}

\subsection{Quantum}

Below we give more precise definitions of locally decodable
codes and related notions, but we first briefly explain the standard 
notation of quantum computing. We refer
to Nielsen and Chuang~\cite{nielsen&chuang:qc} for more details.
A {\em qubit} is a linear combination of the basis states $\ket{0}$ 
and $\ket{1}$, also viewed as a 2-dimensional complex vector:
$$
\alpha_0\ket{0}+\alpha_1\ket{1}=\left(\begin{array}{c}\alpha_0\\ \alpha_1 \end{array}\right),
$$
where $\alpha_0,\alpha_1$ are complex {\em amplitudes}, 
and $|\alpha_0|^2+|\alpha_1|^2=1$. 

The $2^m$ basis states of an $m$-qubit system are the
$m$-fold tensor products of the states $\ket{0}$ and $\ket{1}$.
For example, the basis states of a 2-qubit system are the four 4-dimensional 
unit vectors $\ket{0}\otimes\ket{0}$, $\ket{0}\otimes\ket{1}$,
$\ket{1}\otimes\ket{0}$, and $\ket{1}\otimes\ket{1}$.
We abbreviate, e.g., $\ket{1}\otimes\ket{0}$ to $\ket{0}\ket{1}$, or 
$\ket{1,0}$, or $\ket{10}$, or even $\ket{2}$ (since 2 is 10 in binary).
With these basis states, an $m$-qubit state $\ket{\phi}$ 
is a $2^m$-dimensional complex unit vector
$$
\ket{\phi}=\sum_{i\in\01^m}\alpha_i\ket{i}.
$$
We use $\bra{\phi}=\ket{\phi}^*$ to denote the conjugate transpose
of the vector $\ket{\phi}$, and $\inp{\phi}{\psi}=\bra{\phi}\cdot\ket{\psi}$
for the inner product between states $\ket{\phi}$ and $\ket{\psi}$.
These two states are {\em orthogonal} if $\inp{\phi}{\psi}=0$.
The {\em density matrix\/} corresponding to $\ket{\phi}$ is the
outer product $\ketbra{\phi}{\phi}$. The density matrix corresponding
to a {\em mixed state\/}, which is in pure state $\ket{\phi_i}$
with probability $p_i$, is $\rho=\sum_i p_i\ketbra{\phi_i}{\phi_i}$.
If a 2-register quantum state has the form 
$\ket{\phi}=\sum_i \sqrt{p_i}\ket{i}\ket{\phi_i}$,
then the state of a system holding only the second register
of $\ket{\phi}$ is described by the (reduced) density matrix 
$\sum_i p_i\ketbra{\phi_i}{\phi_i}$. 

The most general measurement allowed by quantum mechanics is 
a so-called {\em positive operator-valued measurement (POVM)}. 
A $k$-outcome POVM is specified by positive operators 
$E_i=M_i^*M_i$, $1\leq i\leq k$, subject to the condition
that $\sum_i E_i=I$. Given a state $\rho$, the probability of getting
the $i$th outcome is $p_i=\Tr(E_i\rho)=\Tr(M_i\rho M_i^*)$. 
If the outcome is indeed $i$, then the resulting state 
is $M_i\rho M_i^*/\Tr(M_i\rho M_i^*)$.
In particular, if $\rho=\ketbra{\phi}{\phi}$, 
then $p_i=\bra{\phi}E_i\ket{\phi}=\norm{M_i\ket{\phi}}^2$,
and the resulting state is $M_i\ket{\phi}/\norm{M_i\ket{\phi}}$.
A special case is where $k=2^m$ and $B=\{\ket{\psi_i}\}$ forms an
orthonormal basis of the $m$-qubit space.
``Measuring in the $B$-basis'' means that we apply the 
POVM given by $E_i=M_i=\ketbra{\psi_i}{\psi_i}$.
Applying this to a pure state $\ket{\phi}$ gives resulting state 
$\ket{\psi_i}$ with probability $p_i=|\inp{\phi}{\psi_i}|^2$.
Apart from measurements, the basic operations that quantum mechanics 
allows us to do, are {\em unitary\/} (i.e.,\ linear norm-preserving) 
transformations of the vector of amplitudes. 

Finally, a word about quantum {\em queries}.
A query to an $m$-bit string $y$ is commonly formalized as the following
unitary transformation, where $j\in[m]$,
and $b\in\01$ is called the {\em target bit}:
$$
\ket{j}\ket{b}\mapsto\ket{j}\ket{b\oplus y_j}.
$$
A quantum computer may apply this to any superposition.
An equivalent formalization that we will be using here, is:
$$
\ket{c}\ket{j}\mapsto(-1)^{c\cdot y_j}\ket{c}\ket{j}.
$$
Here $c$ is a {\em control bit} that controls 
whether the phase $(-1)^{y_j}$ is added or not.
Given some extra workspace, one query of either type can 
be simulated exactly by one query of the other type.

\subsection{Codes}\label{ssecldcdef}

Below, by a `decoding algorithm' we mean an algorithm (quantum
or classical depending on context) with oracle access to the bits of 
some (possibly corrupted) codeword $y$ for $x$.  
The algorithm gets input $i$ and is supposed to recover $x_i$ 
while making only few queries to $y$.

\begin{definition}
$C:\01^n\rightarrow\01^m$ is a {\em $(q,\delta,\eps)$-locally decodable code}
(LDC) if there is a classical randomized decoding algorithm $A$ such that 
\begin{enumerate}
\item $A$ makes at most $q$ queries to $y$, non-adaptively.
\item For all $x$ and $i$, and all $y\in\01^m$ with Hamming distance
$d(C(x),y)\leq\delta m$ we have $\Pr[A^y(i)=x_i]\geq 1/2+\eps$.
\end{enumerate}
The LDC is called {\em linear\/} if $C$ is a linear function over $GF(2)$
(i.e., $C(x+y)=C(x)+C(y)$).

By allowing $A$ to be a quantum computer and to make queries in superposition,
we can similarly define $(q,\delta,\eps)$-locally {\em quantum}-decodable 
codes (LQDCs).
\end{definition}

It will be convenient to work with {\em non-adaptive\/} queries,
as used in the above definition, so the distribution on the queries 
that $A$ makes is independent of $y$.
However, our main lower bound also holds for adaptive queries,
see the first remark at the end of Section~\ref{ssecldclower}.

\subsection{Private information retrieval}

Next we formally define private information retrieval schemes.

\begin{definition}
A one-round, $(1-\delta)$-secure, $k$-server private information retrieval
(PIR) scheme with recovery probability $1/2+\eps$, query size $t$, 
and answer size $a$, consists of a randomized algorithm representing the user,
and $k$ deterministic algorithms $S_1,\ldots,S_k$ (the servers), such that
\begin{enumerate}
\item 
On input $i\in[n]$, the user produces $k$ $t$-bit queries 
$q_1,\ldots,q_k$ and sends these to the respective servers. 
The $j$th server sends back an $a$-bit string $a_j=S_j(x,q_j)$. 
The user outputs a bit $b$ depending on $i,a_1,\ldots,a_k,$ and his randomness.
\item
For all $x$ and $i$, the probability (over the user's randomness) 
that $b=x_i$ is at least $1/2+\eps$. 
\item
For all $x$ and $j$, the distributions on $q_j$ (over the user's randomness)
are $\delta$-close (in total variation distance) for different $i$.
\end{enumerate}
The scheme is called {\em linear\/} if, for every $j$ and $q_j$, 
the $j$th server's answer $S_j(x,q_j)$ is a linear combination 
(over $GF(2)$) of the bits of $x$.
\end{definition}
All known upper bounds on PIR have one round, $\eps=1/2$ (perfect recovery) 
and $\delta=0$ (the servers get no information whatsoever about $i$).
Below we will assume one round and $\delta=0$ 
without mentioning this further.
We can straightforwardly generalize these definitions 
to {\em quantum\/} PIR for the case where $\delta=0$ 
(the server's state after the query should be independent of $i$), 
and that is the only case we will need here.

\section{Lower Bound for Locally Decodable Codes with Two Queries}
 
The proof has two parts, each with a clear intuition 
but requiring quite a few technicalities:
\begin{enumerate}
\item A 2-query LDC gives a 1-query LQDC, because one quantum query
can compute the same Boolean functions as two classical queries 
(albeit with slightly worse error probability).
\item The length $m$ of a 1-query LQDC must be exponential, because
a uniform superposition over all its indices turns out to be a $\log m$-qubit 
{\em quantum random access code\/} for $x$, for which 
a linear lower bound is already known~\cite{nayak:qfa}.
\end{enumerate}

\subsection{From 2 classical queries to 1 quantum query}

The key to the first step is the following lemma:

\begin{lemma}\label{lem2clqueries}
Let $f:\01^2\rightarrow\01$ and suppose we can make queries to the
bits of some input string $a=a_1a_2\in\01^2$.  
There exists a quantum algorithm that makes only one query 
(one that is independent of $f$) and outputs $f(a)$ with probability 
{\em exactly\/} $11/14$, and outputs $1-f(a)$ otherwise.
\end{lemma}

\begin{proof}
The quantum algorithm makes the query
$
\frac{1}{\sqrt{3}}\left(\ket{0}\ket{1}+\ket{1}\ket{1}+\ket{1}\ket{2}\right),
$
where the first bit is the control bit, and the appropriate phase
$(-1)^{a_j}$ is added in front of $\ket{j}$ if the control bit is 1.
The result of the query is the state
$$
\ket{\phi}=\frac{1}{\sqrt{3}}\left(\ket{0}\ket{1}+
(-1)^{a_1}\ket{1}\ket{1}+(-1)^{a_2}\ket{1}\ket{2}\right).
$$
The algorithm then measures this state in a basis containing 
the following four states ($b\in\01^2$):
$$
\ket{\psi_b}=\frac{1}{2}\left(\ket{0}\ket{1}+(-1)^{b_1}\ket{1}\ket{1}+
(-1)^{b_2}\ket{1}\ket{2}+(-1)^{b_1+b_2}\ket{0}\ket{2}\right).
$$
Note that these four states are orthogonal to each other.

The probability of getting outcome $a$ is $|\inp{\phi}{\psi_a}|^2=3/4$,
and each of the other 3 outcomes has probability $1/12$.
The algorithm determines its output based on $f$ and 
on the measurement outcome $b$.
We distinguish 3 cases for $f$:
\begin{enumerate}
\item $|f(1)^{-1}|=1$ (the case $|f(1)^{-1}|=3$ is completely analogous,
with 0 and 1 reversed). 
If $f(b)=1$, then the algorithm outputs 1 with probability 1. 
If $f(b)=0$ then it outputs 0 with probability $6/7$ and 1 with probability
$1/7$.
Accordingly, if $f(a)=1$, then the probability of outputting 1 is
$\Pr[f(b)=1]\cdot 1+\Pr[f(b)=0]\cdot 1/7=3/4+1/28=11/14.$
If $f(a)=0$, then the probability of outputting 0 is
$\Pr[f(b)=0]\cdot 6/7=(11/12)\cdot(6/7)=11/14.$
\item $|f(1)^{-1}|=2$. Then $\Pr[f(a)=f(b)]=3/4+1/12=5/6$.
If the algorithm outputs $f(b)$ with probability $13/14$
and outputs $1-f(b)$ with probability $1/14$, 
then its probability of outputting $f(a)$ is exactly $11/14$.
\item $f$ is constant. In that case the algorithm just outputs 
that value with probability $11/14$.
\end{enumerate}
\vspace*{-2em}
\end{proof}

Peter H\o yer (personal communication) recently improved 
the $11/14$ in the above lemma to $9/10$, which we can show 
to be optimal.  Using our lemma we can prove:

\begin{theorem}\label{thclasqldc}
A $(2,\delta,\eps)$-locally decodable code is 
a $(1,\delta,4\eps/7)$-locally quantum-decodable code.
\end{theorem}

\begin{proof}
Consider some $i$, $x$, and $y$ such that $d(C(x),y)\leq\delta m$.
The 1-query quantum decoder will use the same randomness as the 
2-query classical decoder. The random string of the classical decoder
determines two indices $j,k\in[m]$ and an $f:\01^2\rightarrow\01$ such that
$$
\Pr[f(y_j,y_k)=x_i]=p\geq 1/2+\eps,
$$
where the probability is taken over the decoder's randomness.
We now use Lemma~\ref{lem2clqueries} to obtain a 1-query quantum decoder
that outputs some bit $o$ such that
$$
\Pr[o=f(y_j,y_k)]=11/14.
$$
The success probability of this quantum decoder is:%
\footnote{Here we use the `exactly' part of Lemma~\ref{lem2clqueries}. 
To see what could go wrong if the `exactly' were `at least', suppose 
the classical decoder outputs ${\rm AND}(y_1,y_2)=x_i$ with probability $3/5$ 
and ${\rm XOR}(y_3,y_4)=1-x_i$ with probability $2/5$. 
Then it outputs $x_i$ with probability $3/5>1/2$.
However, if our quantum procedure computes ${\rm AND}(y_1,y_2)$ with 
success probability $11/14$ but ${\rm XOR}(y_3,y_4)$ with success 
probability 1, then its recovery probability is $(3/5)(11/14)<1/2$.}
\begin{eqnarray*}
\Pr[o=x_i] & = & \Pr[o=f(y_j,y_k)]\cdot\Pr[f(y_j,y_k)=x_i]+
\Pr[o\neq f(y_j,y_k)]\cdot\Pr[f(y_j,y_k)\neq x_i]\\
 & = & \frac{11}{14}p+\frac{3}{14}(1-p)
 \ = \ \frac{3}{14}+\frac{4}{7}p
 \ \geq \ \frac{1}{2}+\frac{4\eps}{7}.
\end{eqnarray*}
\vspace*{-3em}

\end{proof}

\subsection{Exponential lower bound for 1-query LQDCs}

A quantum {\em random access code} is an encoding 
$x\mapsto\rho_x$ of $n$-bit strings $x$ into $m$-qubit states $\rho_x$, 
such that any bit $x_i$ can be recovered with some probability $p\geq 1/2+\eps$
from $\rho_x$. 
The following lower bound is known on the length of such 
quantum codes~\cite{nayak:qfa} (see Appendix~\ref{appqrac} for a proof).

\begin{theorem}[Nayak]\label{thnayakqrac}
An encoding $x\mapsto\rho_x$ of $n$-bit strings into $m$-qubit states
with recovery probability at least $p$, has $m\geq(1-H(p))n$.
\end{theorem}

This allows us to prove an exponential lower bound for 1-query LQDC:

\begin{theorem}\label{thlower1lqdc}\label{thlqdcbound}
If $C:\01^n\rightarrow\01^m$ is a $(1,\delta,\eps)$-locally 
quantum-decodable code, then 
$$
m\geq 2^{cn-1},
$$
for $c=1-H(1/2+\delta\eps/4)$.
\end{theorem}

\begin{proof}
We fix $i$.
Let $\ket{Q}=\sum_{c\in\01,j\in[m]}\alpha_{cj}\ket{c}\ket{j}$ be the 
query that the quantum decoder makes to recover $x_i$. 
Let $D$ and $I-D$ be the two POVM operators that the decoder uses
on the state $\ket{R}$ returned by the query, corresponding 
to outcomes 1 and 0, respectively. 
Its probability of outputting 1 on $\ket{R}$ is 
$p(R)=\bra{R}D\ket{R}=\norm{\sqrt{D}\ket{R}}^2$.
Without loss of generality, we assume that
all $\alpha_{cj}$ are non-negative reals 
(this is the most general query a quantum decoder can ask,
because complex phases and entanglement with its workspace can always 
be added by the decoder {\em after\/} the query).
Since $C$ is a LQDC,
the decoder can recover $x_i$ with probability $1/2+\eps$ from the state
$$
\sum_{c\in\01,j\in[m]}\alpha_{cj}(-1)^{c\cdot y_j}\ket{c}\ket{j}
$$ 
for every $y$ such that $d(C(x),y)\leq\delta m$.
Our goal below is to show that we can also recover $x_i$ 
with probability $1/2+\delta\eps/4$ from the uniform state
$$
\ket{U(x)}=\frac{1}{\sqrt{2m}}\sum_{c\in\01,j\in[m]}(-1)^{c\cdot C(x)_j}\ket{c}\ket{j}.
$$
Since $\ket{U(x)}$ is independent of $i$, we can actually
recover any bit $x_j$ with that probability. Hence $\ket{U(x)}$ 
is a $(\log(m)+1)$-qubit random access code for $x$.
Applying Theorem~\ref{thnayakqrac} gives the result.

Inspired by the ``smoothing'' technique of~\cite{katz&trevisan:ldc},
we split the amplitudes $\alpha_j$ 
of the query $\ket{Q}$ 
into small and large ones:
$A=\{cj:\alpha_{cj}\leq\sqrt{1/\delta m}\}$ and 
$B=\{cj:\alpha_{cj}>\sqrt{1/\delta m}\}$.
Since the query does not affect the $\ket{0}\ket{j}$-states,
we can assume without loss of generality
that $\alpha_{0j}$ is the same for all $j$,
so $\alpha_{0j}\leq 1/\sqrt{m}\leq 1/\sqrt{\delta m}$ and hence $0j\in A$.
Let $a=\sqrt{\sum_{cj\in A}\alpha_{cj}^2}$ be the norm of the 
``small-amplitude'' part.
Since $\sum_{cj\in B}\alpha_{cj}^2\leq 1$, we have $|B|<\delta m$.
Define non-normalized states
$$
\ket{A(x)} =  \displaystyle\sum_{cj\in A}(-1)^{c\cdot C(x)_j}\alpha_{cj}\ket{c}\ket{j} \mbox{ \ \ and \ \ }
\ket{B} = \displaystyle\sum_{cj\in B}\alpha_{cj}\ket{c}\ket{j}.
$$
The states $\ket{A(x)}+\ket{B}$ and $\ket{A(x)}-\ket{B}$ each
correspond to a $y\in\01^m$ that is corrupted (compared to $C(x)$)
in at most $|B|\leq\delta m$ positions, so the decoder can recover 
$x_i$ from each of these states. If $x$ has $x_i=1$, then
$$
p(A(x)+B)\geq 1/2+\eps\mbox{ \ and \ }
p(A(x)-B)\geq 1/2+\eps.
$$ 
Since $p(A\pm B)=p(A)+p(B)\pm(\bra{A}D\ket{B}+\bra{B}D\ket{A})$, 
averaging the previous two inequalities gives
$$
p(A(x))+p(B)\geq 1/2+\eps.
$$
Similarly, if $x'$ has $x'_i=0$, then 
$$
p(A(x'))+p(B)\leq 1/2-\eps.
$$
Hence, for the normalized states $\ket{A(x)}/a$ and $\ket{A(x')}/a$ we have
$$
p(A(x)/a)-p(A(x')/a)\geq 2\eps/a^2.
$$
Since this holds for every $x,x'$ with $x_i=1$ and $x'_i=0$,
there are constants $q_1,q_0\in[0,1]$, $q_1-q_0\geq 2\eps/a^2$, 
such that $p(A(x)/a)\geq q_1$ whenever $x_i=1$ 
and $p(A(x)/a)\leq q_0$ whenever $x_i=0$.

If we had a copy of the state $\ket{A(x)}/a$,
then we could run the procedure below to recover $x_i$. 
Here we assume that $q_1\geq 1/2+\eps/a^2$ (if not, then we must have
$q_0\leq 1/2-\eps/a^2$ and we can use the same argument with 
0 and 1 reversed),  
and that $q_1+q_0\geq 1$ (if not, then
$q_0\leq 1/2-\eps/a^2$ and we're already done).
\begin{quote}
Output 0 with probability $q=1-1/(q_1+q_0)$,\\
and otherwise output the result of running the decoder's POVM 
on $\ket{A(x)}/a$.
\end{quote}
If $x_i=1$, then the probability that this procedure outputs 1 is 
$$
(1-q)p(A(x)/a)\geq (1-q)q_1=\frac{q_1}{q_1+q_0}=
\frac{1}{2}+\frac{q_1-q_0}{2(q_1+q_0)}\geq \frac{1}{2}+\frac{\eps}{2a^2}.
$$
If $x_i=0$, then the probability that it outputs 0 is
$$
q+(1-q)(1-p(A(x)/a))\geq
q+(1-q)(1-q_0)=1-\frac{q_0}{q_1+q_0}=\frac{q_1}{q_1+q_0}\geq 
\frac{1}{2}+\frac{\eps}{2a^2}.
$$
Thus, we can recover $x_i$ with good probability if we had the state
$\ket{A(x)}/a$.

It remains to show how we can obtain $\ket{A(x)}/a$ from 
$\ket{U(x)}$ with reasonable probability. This we do by applying a POVM 
with operators $M^\dagger M$ and $I-M^\dagger M$ to $\ket{U(x)}$, where
$M=\sqrt{\delta m}\sum_{cj\in A}\alpha_{cj}\ketbra{cj}{cj}$.
Both $M^\dagger M$ and $I-M^\dagger M$ are positive
operators (as required for a POVM) 
because $0\leq \sqrt{\delta m}\alpha_{cj}\leq 1$ for all $cj\in A$.
The POVM gives the first outcome with probability 
$$
\bra{U(x)}M^\dagger M\ket{U(x)}=
\frac{\delta m}{2m}\sum_{cj\in A}\alpha_{cj}^2=\delta a^2/2.
$$
In this case we have obtained the normalized version of $M\ket{U(x)}$,
which is $\ket{A(x)}/a$, so then we can run 
the above procedure to recover $x_i$.
If the measurement gives the second outcome, 
then we just output a fair coin flip. 
Thus we recover $x_i$ from $\ket{U(x)}$ with probability at least
$$
(\delta a^2/2)(1/2+\eps/2a^2)+(1-\delta a^2/2)1/2=1/2+\delta\eps/4.
$$
\vspace*{-3em}

\end{proof}

\subsection{Exponential lower bound for 2-query LDCs}\label{ssecldclower}

\begin{theorem}\label{mainthm}
If $C:\01^n\rightarrow\01^m$ is a $(2,\delta,\eps)$-locally 
decodable code, then 
$$
m\geq 2^{cn-1},
$$
for $c=1-H(1/2+3\delta\eps/14)$.
\end{theorem}

\begin{proof}
The theorem follows by combining Theorems~\ref{thclasqldc} 
and~\ref{thlower1lqdc}.
Straightforwardly, this would give a constant of $1-H(1/2+\delta\eps/7)$.
We get the better constant claimed here by observing that the 1-query
LQDC derived from the 2-query LDC actually has $1/3$ of the overall
squared amplitude on queries where the control bit $c$ is zero
(and all those $\alpha_{0j}$ are in $A$). 
Hence in the proof of Theorem~\ref{thlower1lqdc},
we can redefine ``small amplitude'' to $\alpha_{cj}\leq\sqrt{2/3\delta m}$,
and still $B$ will have at most $\delta m$ elements because
$\sum_{cj\in B}\alpha_{cj}^2\leq 2/3$. This in turns allows
us to make $M$ a factor $\sqrt{3/2}$ larger, which improves 
the probability of getting $\ket{A(x)}/a$ from $\ket{U(x)}$ to
$3\delta a^2/4$ and the recovery probability to $1/2+3\delta\eps/8$.
Combining that with Theorem~\ref{thclasqldc} (which makes $\eps$ a factor 
$4/7$ smaller) gives $c=1-H(1/2+3\delta\eps/14)$, as claimed.
\end{proof}

\noindent
{\bf Remarks:}

{\bf (1)} 
A $(2,\delta,\eps)$-LDC with {\em adaptive} queries gives a
$(2,\delta,\eps/2)$-LDC with non-adaptive queries: if query $q_1$ would
be followed by query $q_2^0$ or $q_2^1$ depending on the outcome of $q_1$,
then we can just guess in advance whether to query $q_1$ and $q_2^0$,
or $q_1$ and $q_2^1$. With probability 1/2, the second query will
be the one we would have made in the adaptive case and we're fine,
in the other case we just flip a coin, giving overall recovery 
probability $1/2(1/2+\eps)+1/2(1/2)=1/2+\eps/2$.
Thus we also get slightly weaker but still exponential lower bounds 
for {\em adaptive} 2-query LDCs.

{\bf (2)}
For a $(2,\delta,\eps)$-LDC where the decoder's output is the XOR
of its two queries, we can give a better reduction
than in Theorem~\ref{thclasqldc}. In this case, the quantum decoder 
can apply his query to
$
\frac{1}{\sqrt{2}}\left(\ket{1}\ket{1}+\ket{1}\ket{2}\right),
$
giving
$$
\frac{1}{\sqrt{2}}\left((-1)^{a_1}\ket{1}\ket{1}+(-1)^{a_2}\ket{1}\ket{2}\right)=
(-1)^{a_1}\frac{1}{\sqrt{2}}\left(\ket{1}\ket{1}+(-1)^{a_1\oplus a_2}\ket{1}\ket{2}\right),
$$
and extract $a_1\oplus a_2$ from this with certainty. 
Thus the recovery probability remains $1/2+\eps$
instead of going down to $1/2+4\eps/7$.
Accordingly, we also get slightly better lower bounds for 2-query LDCs 
where the output is the XOR of the two queried bits, 
namely $c=1-H(1/2+3\delta\eps/8)$.

{\bf (3)}
In Appendix~\ref{applarge} we extend the lower bound to larger alphabets.

\section{Locally Quantum-Decodable Codes with Few Queries}

The second remark of Section~\ref{ssecldclower} immediately generalizes to:

\begin{theorem}\label{thldctolqdc}
A $(2q,\delta,\epsilon)$-LDC where the decoder's output is 
the XOR of the $2q$ queried bits, is a $(q,\delta,\eps)$-LQDC.
\end{theorem}

LDCs with $q$ queries can be obtained from $q$-server PIR schemes 
with 1-bit answers by concatenating the answers that the servers 
give to all possible queries of the user.
Beimel et al.~\cite[Corollary~4.3]{bikr:improvedpir} recently improved 
the best known upper bounds on $q$-query LDCs, based on their improved 
PIR construction. 
They give a general upper bound $m=2^{n^{O(\log\log q/q\log q)}}$ 
for $q$-query LDCs, for some constant depending on $\delta$ and $\epsilon$,
as well as more precise estimates for small $q$. 
In particular, for $q=4$ they construct an LDC of length $m=2^{O(n^{3/10})}$.
All their LDCs are of the XOR-type, so we can reduce the number of
queries by half when allowing quantum decoding.
For instance, their 4-query LDC is a 2-query LQDC with 
length $m=2^{O(n^{3/10})}$. In contrast, any 2-query LDC requires 
length $m=2^{\Omega(n)}$ as we proved above.

For general LDCs we can do something nearly as good,
using van Dam's result that a $q$-bit oracle can be recovered with probability 
nearly 1 using $q/2+O(\sqrt{q})$ quantum queries~\cite{dam:oracle}:

\begin{theorem}
A $(q,\delta,\epsilon)$-LDC is a $(q/2+O(\sqrt{q}),\delta,\eps/2)$-LQDC.
\end{theorem}

\section{Private Information Retrieval}

\subsection{Lower bounds for classical PIR}

As mentioned, there is a close connection between locally 
decodable codes and private information retrieval.
Our techniques allow us to give new lower bounds for 2-server
PIRs.  Again we give
a 2-step proof: a reduction of 2 classical servers to 1 quantum
server, combined with a lower bound for 1-server quantum PIR.

\begin{theorem}\label{th2to1pir}
If there exists a classical 2-server PIR scheme with $t$-bit queries,
1-bit answers, and recovery probability $1/2+\eps$, then
there exists a quantum 1-server PIR scheme with $(t+2)$-qubit queries,
$(t+2)$-qubit answers, and recovery probability $1/2+4\eps/7$.
\end{theorem}

\begin{proof}
The proof is analogous to the proof for locally decodable codes.
If we let the quantum user use the same randomness as the classical one,
the problem boils down 
to computing some $f(a_1,a_2)$, where $a_1$ is the first server's 1-bit answer 
to query $q_1$, and $a_2$ is the second server's 1-bit answer to query $q_2$.
However, in addition we now have to hide $i$ from the quantum server.
This we do by making the quantum user set up the $(4+t)$-qubit state
$$
\frac{1}{\sqrt{3}}\left(\ket{0}\ket{0,0^t}+\ket{1}\ket{1,q_1}+\ket{2}\ket{2,q_2}\right),
$$
where `$0^t$' is a string of $t$ 0s.
The user sends everything but the first register to the server.
The state of the server is now a uniform mixture of $\ket{0,0^t}$,
$\ket{1,q_1}$, and $\ket{2,q_2}$. By the security of the classical
protocol, $\ket{1,q_1}$ contains no information about $i$ (averaged
over the user's randomness), and the same holds for $\ket{2,q_2}$.
Hence the server gets no information about $i$.

The quantum server then puts $(-1)^{a_j}$ in front of
$\ket{j,q_j}$ ($j\in\{1,2\}$), leaves $\ket{0,0^t}$ alone, 
and sends everything back.
Note that we need to supply the name of the classical server 
$j\in\{1,2\}$ to tell the server in superposition whether 
it should play the role of server 1 or~2. The user now has 
$$
\frac{1}{\sqrt{3}}\left(\ket{0}\ket{0,0^t}+(-1)^{a_1}\ket{1}\ket{1,q_1}+
(-1)^{a_2}\ket{2}\ket{2,q_2}\right).
$$
{}From this we can compute $f(a_1,a_2)$ with success probability exactly
$11/14$, giving overall recovery probability $1/2+4\eps/7$ as before.
\end{proof}

Combining the above reduction with the quantum random access code
lower bound, we obtain the first $\Omega(n)$ lower bound that 
holds for all 1-bit-answer 2-server PIRs, not just for linear ones.

\begin{theorem}\label{thpirlower}
A classical 2-server PIR scheme with $t$-bit queries,
1-bit answers, and recovery probability $1/2+\eps$, has
$t\geq(1-H(1/2+4\eps/7))n-2$.
\end{theorem}

\begin{proof}
We first reduce the 2 classical servers to 1 quantum server
in the way of Theorem~\ref{th2to1pir}.
Now consider the state of the quantum PIR scheme after the user
sends his $(t+2)$-qubit message:
$$
\ket{\phi_i}=\sum_r\sqrt{p_r}\ket{r}\frac{1}{\sqrt{3}}
\left(\ket{0}\ket{0,0^t}+\ket{1}\ket{1,q_1(r,i)}+\ket{2}\ket{2,q_2(r,i)}\right).
$$
Here the $p_r$ are the classical probabilities of the user 
(these depend on $i$) and $q_j(r,i)$ is the $t$-bit query that
the user sends to server $j$ in the classical 2-server scheme,
if he wants $x_i$ and has random string $r$.
Letting $B=\{0^{t+1}\}\cup\{1,2\}\times\{0,1\}^t$ be the server's 
basis states, we can write $\ket{\phi_i}$ as:
$$
\ket{\phi_i}=\sum_{b\in B}\lambda_b\ket{a_{ib}}\ket{b}.
$$
Here the $\ket{a_{ib}}$ are pure states that do not depend on $x$. 
The coefficients $\lambda_b$ are non-negative reals 
that do not depend on $i$, for otherwise a measurement
of $b$ would give the server information about $i$ (contradicting privacy).
The server then tags on the appropriate phase $s_{bx}$, which is 1 for
$b=0^{t+1}$ and $(-1)^{S_j(x,q_j)}$ for $b=jq_j$, $j\in\{1,2\}$.
This gives
$$
\ket{\phi_{ix}}=\sum_{b\in B}\lambda_b\ket{a_{ib}}s_{bx}\ket{b}.
$$
Now the following pure state will be a random access code for $x$
$$
\ket{\psi_{x}}=\sum_{b\in B}\lambda_b s_{bx}\ket{b},
$$
because a user can unitarily map
$\ket{0}\ket{b}\mapsto\ket{a_{ib}}\ket{b}$ to map
$\ket{0}\ket{\psi_x}\mapsto\ket{\phi_{ix}}$,
from which he can get $x_i$ with probability $p=1/2+4\eps/7$
by completing the quantum PIR protocol.
The state $\ket{\psi_x}$ has $t+2$ qubits,
hence from Theorem~\ref{thnayakqrac} we obtain $t\geq (1-H(p))n-2$.
\end{proof}

\noindent
In Appendix~\ref{applarge} we extend this bound to classical 2-server PIR
schemes with larger answer size.  

For the special case where
the classical PIR outputs the XOR of the two answer bits,
we can improve our lower bound to $t\geq (1-H(1/2+\eps))n-1$.
In particular, $t\geq n-1$ in case of perfect recovery ($\eps=1/2$),
which is tight.
Very recently but independently of our work, 
Beigel, Fortnow, and Gasarch~\cite{bfg:pir} found a classical proof 
that a 2-server PIR with perfect recovery and 1-bit answers 
needs query length $t\geq n-2$ (no matter whether it uses XOR or not).

\subsection{Upper bounds for quantum PIR}

The best known LDCs are derived from classical PIR
schemes with 1-bit answers where the output is the XOR of the 1-bit answers
that the user receives. By allowing quantum queries, we can reduce
the number of queries by half to obtain more efficient LQDCs.
Similarly, we can also turn the underlying classical $k$-server PIR 
schemes directly into quantum PIR schemes with $k/2$ servers. 

Most interestingly, there exists
a 4-server PIR with 1-bit answers and communication complexity
$O(n^{3/10})$~\cite[Example~4.2]{bikr:improvedpir}. 
This gives us a quantum 2-server PIR scheme with
$O(n^{3/10})$ communication, improving upon the communication required 
by the best known classical 2-server PIR scheme, which has been $O(n^{1/3})$
ever since the introduction of PIR by Chor et al.~\cite{cgks:pir}.
In the introduction we mentioned also some upper bounds for $k>2$,
which are obtained similarly.

\subsection*{Acknowledgments}
We thank Luca Trevisan for many insightful comments throughout this
work and also for allowing us to include Lemma~\ref{smooth}
in Appendix~\ref{appqrac}. 
We also thank Harry Buhrman, Peter H\o yer, 
Hartmut Klauck, Ashwin Nayak, Kenji Obata, 
Pranab Sen (and via him also Rahul Jain), 
Mario Szegedy, and Ashish Thapliyal for helpful discussions.
We thank Amos Beimel for sending us a version of~\cite{bikr:improvedpir} 
and Bill Gasarch for sending us a version of~\cite{bfg:pir}.


\appendix

\section{Lower Bound for Quantum Random Access Codes}\label{appqrac}

As mentioned before, a quantum random access code is an encoding 
$x\mapsto\rho_x$, such that any bit $x_i$ can be recovered with some 
probability $p\geq 1/2+\eps$ from $\rho_x$. Below we reprove
Nayak's~\cite{nayak:qfa} linear lower bound on the length $m$ 
of such encodings.

We assume familiarity with the following notions from quantum information 
theory, referring to~\cite[Chapters~11 and~12]{nielsen&chuang:qc} 
for more details. Very briefly, if we have a bipartite quantum system $AB$ 
(given by some density matrix), then we use $A$ and $B$ 
to denote the states (reduced density matrices) of the individual systems.
$S(A)=-\Tr(A\log A)$ is the {\em (Von Neumann) entropy} of $A$;
$S(A|B)=S(AB)-S(B)$ is the {\em conditional entropy} of $A$ given $B$;
and $S(A:B)=S(A)+S(B)-S(AB)=S(A)-S(A|B)$ is the {\em mutual
information} between $A$ and $B$.

We define an $n+m$-qubit state $XM$ as follows:
$$
\frac{1}{2^n}\sum_{x\in\01^n}\ketbra{x}{x}\otimes\rho_x.
$$
We use $X$ to denote the first subsystem, $X_i$ for its individual bits, 
and $M$ for the second subsystem.
By \cite[Theorem~11.8.4]{nielsen&chuang:qc} we have
$$
S(XM)=n+\frac{1}{2^n}\sum_x S(\rho_x)\geq n=S(X).
$$
Since $M$ has $m$ qubits we have $S(M)\leq m$, hence
$$
S(X:M)=S(X)+S(M)-S(XM)\leq S(M)\leq m.
$$
Using a chain rule for relative entropy, 
and the subadditivity of Von Neumann entropy we get
$$
S(X|M)=\sum_{i=1}^n S(X_i|X_1\ldots X_{i-1}M)\leq\sum_{i=1}^n S(X_i|M).
$$
Since we can predict $X_i$ from $M$ with success probability $p$, 
Fano's inequality implies
$$
H(p)\geq S(X_i|M).
$$
In fact, Fano's inequality even applies under the weaker assumption 
that the success probability in predicting $x_i$ is $p$ 
only when {\em averaged\/} over all $x$.
Putting the above equations together we obtain
$$
(1-H(p))n\leq S(X)-\sum_{i=1}^n S(X_i|M)\leq S(X)-S(X|M)=S(X:M)\leq m.
$$

\section{Extension to Larger Alphabets}\label{applarge}

In this section we extend our lower bounds for binary 2-query LDCs 
to the case of larger alphabets (and our bounds for binary 
2-server PIR schemes to the case of larger answers).
For simplicity we assume the alphabet is $\Sigma=\01^\ell$, so a
query to position $j$ now returns an $\ell$-bit string $C(x)_j$.
The definition of $(q,\delta,\eps)$-LDC from Section~\ref{ssecldcdef}
carries over immediately, with $d(C(x),y)$ now measuring the
Hamming distance between $C(x)\in\Sigma^m$ and $y\in\Sigma^m$.

We will need the notion of {\em smooth\/} codes and their 
connection to LDCs as stated in~\cite{katz&trevisan:ldc}. 

\begin{definition}
$C:\01^n\rightarrow\Sigma^m$ is a $(q,c,\epsilon)$-{\em smooth
code} if there is a classical randomized decoding algorithm $A$ such that
\begin{enumerate}
\item $A$ makes at most $q$ queries, non-adaptively.
\item For all $x$ and $i$ we have 
$\Pr[A^{C(x)}(i)=x_i]\geq 1/2+\eps$
\item For all $x$, $i$, and $j$, the probability that on input
$i$ machine $A$ queries index $j$ is at most $c/m$.
\end{enumerate}  
\end{definition}

Note that smooth codes only require good decoding 
on codewords $C(x)$, not on $y$ that are close to $C(x)$.  
Katz and Trevisan~\cite[Theorem~1]{katz&trevisan:ldc} established
the following connection:

\begin{theorem}[Katz and Trevisan]\label{ldctosmooth}
Let $C:\01^n\rightarrow\Sigma^m$ be a $(q,\delta,\eps)$-locally
decodable code. Then $C$ is also a $(q,q/\delta,\eps)$-smooth code.
\end{theorem}

In order to prove the exponential lower bound for LDCs over non-binary
alphabet $\Sigma$, we will reduce a smooth code over $\Sigma$ to a 
somewhat longer {\em binary\/} smooth code that works well
{\em averaged over $x$}. Then, we will show a lower bound on such 
average-case binary smooth codes in a way very similar to the proof of
Theorem~\ref{mainthm}.
The following key lemma was suggested to us by Luca Trevisan.

\begin{lemma}[Trevisan]\label{smooth}
Let $C:\01^n\rightarrow\Sigma^m$ be a $(2,c,\eps)$-smooth code.
Then, there exists a $(2,c\cdot 2^\ell,\eps/2^{2\ell})$-smooth code
$C':\01^n\rightarrow\01^{m\cdot 2^{\ell}}$ that is good {\em on average}, 
i.e., there is a decoder $A$ such that for all $i\in[n]$
$$
{1\over 2^n}\sum_{x\in\01^n}\Pr[A^{C'(x)}(i)=x_i]\geq {1\over
2}+{\eps\over{2^{2\ell}}}. 
$$
\end{lemma}  

\begin{proof}
We form the new binary code $C'$ by replacing each symbol
$C(x)_j\in\Sigma$ of the old code by its Hadamard code, which consists
of $2^\ell$ bits. The length of $C'(x)$ is $m\cdot 2^{\ell}$ bits. 
The new decoding algorithm uses the same randomness as the old one.
Let us fix the two queries $j,k\in[m]$ and the output function 
$f:\Sigma^2\rightarrow\01$ of the old decoder. 
We will describe a new decoding algorithm that is good for an average $x$ 
and looks only at one bit of the Hadamard codes of each of
$a=C(x)_j$ and $b=C(x)_k$. 

First, if for this specific $j,k,f$ we have $\Pr_x[f(a,b)=x_i]\leq 1/2$, 
then the new decoder just outputs a random bit, so in this case it is 
at least as good as the old one for an average $x$. 
Now consider the case $\Pr_x[f(a,b)=x_i]= 1/2+\eta$ for some $\eta>0$.
Switching from the $\{0,1\}$-notation to the $\{-1,1\}$-notation 
enables us to say that $E_x[f(a,b)\cdot x_i]=2\eta$. 
Viewing $a$ and $b$ as two $\ell$-bit strings, we can represent
$f$ by its Fourier representation:
$f(a,b)=\sum_{S,T\subseteq[\ell]}\hat{f}_{S,T}\prod_{s\in S}a_s\prod_{t\in
T} b_t$ and hence 
$$
\sum_{S,T}\hat{f}_{S,T}E_x\left[\prod_{s\in S}a_s\prod_{t\in T} b_t\cdot x_i\right]
=E_x\left[\left(\sum_{S,T}\hat{f}_{S,T}\prod_{s\in S}a_s
\prod_{t\in T}b_t\right)\cdot x_i\right] = E_x[f(a,b)\cdot x_i]=2\eta.
$$
Averaging and using that $|\hat{f}_{S_0,T_0}|\leq 1$,
it follows that there exist subsets $S_0,T_0$ such that 
$$
\left|E_x\left[\prod_{s\in S_0}a_s\prod_{t\in 
T_0} b_t\cdot x_i\right]\right|\geq 
\hat{f}_{S_0,T_0}E_x\left[\prod_{s\in S_0}a_s\prod_{t\in 
T_0} b_t\cdot x_i\right] \geq {2\eta\over 2^{2\ell}}.
$$
Returning to the $\01$-notation, we must have either
$$ 
\Pr_x[S_0\cdot a\oplus T_0\cdot b=x_i]\geq
1/2+\eta/2^{2\ell}
$$
or 
$$
\Pr_x[S_0\cdot a \oplus T_0\cdot b=x_i]\leq 1/2-\eta/2^{2\ell},
$$
where $S_0\cdot a$ and $T_0\cdot b$ denote
inner products mod 2 of $\ell$-bit strings.
Accordingly, either the XOR of the two bits $S_0\cdot a$ and $T_0\cdot b$,
or its negation, predicts $x_i$ with average probability 
$\geq 1/2+\eta/2^{2\ell}$.   
Both of these bits are in the binary code $C'(x)$.
The $c$-smoothness of $C$ translates into $c\cdot 2^{\ell}$-smoothness of $C'$.
Averaging over the classical randomness (i.e.\ the choice of $j,k$,
and $f$) gives the lemma.
\end{proof}

This lemma enables us to modify our proof of Theorem~\ref{mainthm}
so that it works for non-binary alphabets $\Sigma$:

\begin{theorem}\label{thmlarger}
If $C:\01^n\rightarrow\Sigma^m=(\01^\ell)^m$ is a $(2,\delta,\eps)$-locally 
decodable code, then 
$$
m\geq 2^{cn-\ell},
$$
for $c=1-H(1/2+\delta\eps/2^{3\ell+1})$.
\end{theorem}

\begin{proof}
Using Theorem~\ref{ldctosmooth} and Lemma~\ref{smooth}, we turn 
$C$ into a binary $(2,2^{\ell+1}/\delta,\eps/2^{2\ell})$-smooth code $C'$
that has average recovery probability $1/2+\eps/2^{2\ell}$ and
length $m'=m\cdot 2^\ell$ bits. 
Since its decoder XORs its two binary queries, we can reduce this
to one quantum query without any loss in the average recovery probability
(see the second remark following Theorem~\ref{mainthm}).

We now reduce this quantum smooth code to a quantum random access code,
by a modified version of the proof of Theorem~\ref{mainthm}. 
The smoothness of $C'$ implies that all amplitudes $\alpha_j$ 
(which depend on $i$) in the one quantum query satisfy 
$\alpha_j\leq \sqrt{2^{\ell+1}/\delta m'}$.  Hence there is no need
to split the set of $j$'s into $A$ and $B$. Also, the control bit $c$
will always be 1, so we can ignore it. 

Consider $\ket{U(x)}=\frac{1}{\sqrt{m'}}\sum_{j=1}^{m'}(-1)^{C(x)'_j}\ket{j}$,
$\ket{A(x)}=\sum_{j=1}^{m'}\alpha_j(-1)^{C(x)'_j}\ket{j}$,
and POVM operator 
$M=\sqrt{\delta m'/2^{\ell+1}}\sum_j\alpha_j\ket{j}\bra{j}$. 
The probability that the POVM takes us from $\ket{U(x)}$
to $M\ket{U(x)}=\ket{A(x)}$ is now $\bra{U(x)}M^*M\ket{U(x)}=\delta/2^{\ell+1}$.  
Hence $\ket{U(x)}$ forms a random access code with 
average success probability 
$$
p=\frac{\delta}{2^{\ell+1}}\cdot\left(\frac{1}{2}+\frac{\eps}{2^{2\ell}}\right)+
\left(1-\frac{\delta}{2^{\ell+1}}\right)\frac{1}{2}=
\frac{1}{2}+\frac{\delta\eps}{2^{3\ell+1}}.
$$
The $(1-H(p))n$ lower bound for a quantum random access code holds even 
if the recovery probability $p$ is only an average over $x$, 
hence we obtain $\log(m')\geq(1-H(p))n$.   
\end{proof}

We can also extend our linear lower bound on 2-server PIR schemes with answer
length $a=1$ (Theorem~\ref{thpirlower}) to the case of larger answer length.
We use the reduction from PIR to smooth codes given by
Lemma 7.1 of~\cite{gkst:lowerpir}:

\begin{lemma}[GKST]
If there is a classical 2-server PIR scheme with query length $t$,
answer length $a$, and recovery probability $1/2+\eps$,
then there is a $(2,3,\eps)$-smooth code $C:\01^n\rightarrow\Sigma^m$
for $\Sigma=\01^a$ and $m\leq 6\cdot 2^t$.
\end{lemma}

Going through roughly the same steps as for the above LDC lower bound, we get:

\begin{theorem}
A classical 2-server PIR scheme with $t$-bit queries,
$a$-bit answers, and recovery probability $1/2+\eps$, has
$t\geq\Omega(n\eps^2/2^{6a})$.
\end{theorem}

\end{document}